\documentclass[aps,prl,showpacs,twocolumn,groupedaddress,amssymb]{revtex4}

\usepackage{graphicx}
\usepackage{dcolumn}
\usepackage{bm}

\begin{document}

\title{Pondermotive potential  and Backward Raman Scattering in dense quantum plasmas}

\author{S. Son}
\affiliation{18 Caleb Lane, Princeton, NJ, 08540}

\begin{abstract}

The response of dense quantum plasmas in
 the backward Raman scattering  is studied.
The coefficients in the backward Raman scattering 
is found to be underestimated (overestimated) in the classical theory 
if the excited Langmuir wave has  low-wave vector (high-wave vector).
The second order quantum perturbation theory  shows that 
the second harmonic of the pondermotive potential arises naturally even in 
a single particle motion contrary to the classical plasmas. 
\end{abstract}

\maketitle
The study of dense plasmas becomes increasingly important 
in the inertial confinement fusion~\cite{tabak,sonprl,sonpla, sonpla2, sonpla3, sonchain} and the laser-plasma interaction~\cite{cpa, cpa2, cpa4}. 
 In particular, thanks to the great advance in the free electron laser~\cite{freelaser, freelaser2, Free, Free2}, there have been growing interests in 
compressing the x-ray lasers via the BRS in dense plasmas~\cite{sonbackward, malkin1,Fisch}.
Understanding the dense plasmas is challenging 
due to the significant quantum effect; A few  physical processes, 
that deviate significantly 
from the classical prediction, 
have been identified in dense plasmas~\cite{shukla,  sonprl, sonpla, sonlandau, sonbackward, songamma, sonpre, sonpla2, sonpla3}. 
One question, that the author addresses,   is 
how the quantum effects play a role in the BRS. 

In this paper, the quantum mechanical excitation of the Langmuir wave 
in the  BRS is computed, including the electron diffraction and degeneracy. 
To the author's knowledge, this is the first attempt to estimate the quantum 
effects on the BRS in the warm dense matters relevant to the ICF. 
Our study is based on the random phase approximation~\cite{lindhard} and 
the density functional approach. 
The strength of the BRS is overestimated (underestimated)  
by the classical prediction up to 100 percents 
when the wave-vector of the Langmuir wave (pondermotive potential) is comparable (low) to the Fermi wave vector. 
Performing the quantum perturbation theory to the second order, 
The second harmonic of the pondermotive potential is obtained,
which arises  quantum mechanically even in a single particle motion
 contrary to the classical plasmas.

To begin with, 
consider an electron in the presence of the intense counter-propagating x-rays.
The Schroedinger's equation is given as 
\begin{equation} 
i\hbar\frac{\partial \psi}{\partial t}  
= \left[ \frac{1}{2m_e} \left( i\hbar \nabla + \frac{e}{c} \mathbf{A}(x,t) \right)^2 \right] \psi  \mathrm{,} \label{eq:ham}
\end{equation}
where 
\begin{equation}
\frac{e  \mathbf{A}}{ c}  = m_e \left(v^p_{osc} (x,t) 
+ v^s_{osc} (x,t)  \right) \mathrm{,}
\end{equation}
where $v^p_{osc} = e E^p/m_e \omega_1 \cos(k_1 x -\omega_1 t) $ ($v^s_{osc} = e E^s/m_e \omega_2 \cos(k_1 x -\omega_1 t)$) 
is the oscillating velocity of the electron by the pump (seed) x-ray laser,
   $E^p = \omega A^p/c$ ($E^S = \omega A^S /c$), the electric field of the x-ray laser is in the y-direction, the lasers are counter-propagating in the x-direction and we assume that the dispersion relation of the x-rays is the same as vacuum ($\omega_1 = c k_1$, $\omega_2 = c k_2$). 
The Hamiltonian in Eq.~(\ref{eq:ham}) can be expanded as 

\begin{eqnarray}
H = &+&\frac{1}{2m_e}
 \left[ -\hbar^2 \nabla^2 + i m_e \hbar \cdot \nabla \left(v^p_{osc}(x,t) 
+ v^s_{osc}(x,t)  \right) \right] \nonumber \\ &+& \frac{m_e}{2}  \left[v^p_{osc}(x,t) 
+ v^s_{osc}(x,t)  \right]^2    \mathrm{.} \label{eq:ham2}
\end{eqnarray} 
In the regime of our interest, 
 $c \gg v_{osc} $ and $c\gg v_{te} $, where $v_{te}$ is the electron thermal velocity. 
In the absence of the oscillating field $A^p$ and $A^s$, the solution of Eq.~(\ref{eq:ham}) is the plane wave given as $\psi_k = \exp(-i(\hbar k^2/2m_e)t + i k x) $.   
 Since   $\omega_1 = c k_1$ and  $\omega_2 = c k_2$, 
the second term of the right side in Eq.~(\ref{eq:ham}) is very fast frequency if $c \gg v_{osc} $ and $c\gg v_{te} $ and can be ignored. 
The third term has the fast frequency and the slow frequency. The term with the slow frequency is the beating term given as 

\begin{equation} 
V_p =  \frac{m_e}{2} v^p_{osc}v^s_{osc} \cos\left( (k_1+k_2)x - (\omega_1 - \omega_2)t \right)  \mathrm{.} \label{eq:pond} 
\end{equation}
Eq.~(\ref{eq:pond}) is the so-called the pondermotive potential. 

Using the dielectric function approach, the density perturbation from the potential $V_p$ is given as 

\begin{equation}
 \delta n_e =\frac{ \chi_e(k_1 + k_2, \omega_1 - \omega_2) V_p } { \epsilon(k_1 + k_2, \omega_1 - \omega_2)} \mathrm{,} \label{eq:density}  
\end{equation}
where $ \epsilon = 1+ (4 \pi e^2/ k^2) \chi_e $ is the dielectric function. 
The susceptibility  $\chi_e$ can be obtained through the Vlasov theory for the classical plasmas and the random phase approximation for the degenerate plasma~\cite{lindhard}. 
 In classical plasmas, the susceptibility is given as 

\begin{equation}
 \chi_e^C(k, \omega) = \frac{n_e}{m_e} \int \left[ \frac{ \mathbf{k} \cdot \mathbf{\nabla}_v f_e }{\omega - \mathbf{k} \cdot \mathbf{v} }\right]d^3 \mathbf{v} 
\label{eq:cla} 
\end{equation} 
where $m_e$ ($n_e$) is the particle mass (density) and $f_e $ is the distribution with the normalization $\int f_e d^3 \mathbf{v} = 1$.  
In the completely degenerate plasma, the susceptibility is given as

\begin{figure}
\scalebox{0.3}{
\includegraphics[width=1.7\columnwidth, angle=270]{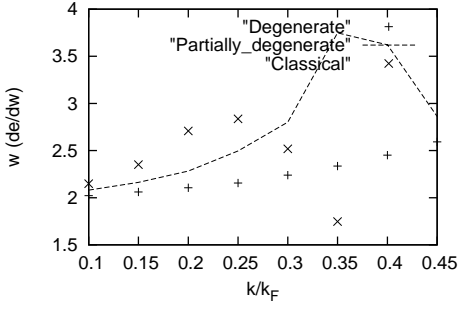}}
\caption{\label{fig1}
$\omega (\partial \epsilon / \partial \omega) $ as a function of the wave vector $k/k_F$ for the classical plasmas, the partially degenerate plasma and completely degenerate plasma. The x-axis is $k / k_F $ and the y-axis is $\omega (\partial \epsilon / \partial \omega) $. 
In this example, $n_e = 10^{24} / \mathrm{cc} $, 
$T_e = E_F = 36\  \mathrm{eV} $.  
}
\end{figure}

\begin{equation}
\chi_e^Q(\mathbf{k},\omega) = \frac{3 n_e}{m_e v_F^2} h(z, u) \mathrm{,} \nonumber \label{eq:quan}\end{equation}    
where $v_F = \sqrt{2E_F/m_e}$ is the Fermi velocity, $E_F = \hbar^2 k_F^2 / 2 m_e $ ($k_F = (3 \pi^2 n_e)^{1/3}$ is the Fermi energy (Fermi wave vector), 
$ z = k / 2 k_F$, $u = \omega /k v_F$, and $h= h_r + i h_i$. 
Note that The Fermi energy is given as
$ E_F =  36.4 \times (n/n_{24})^{2/3} \mathrm{eV}$ where  $n_{24} = 10^{24} / \mathrm{cc}$.
The real part of $h$ is given as 
\begin{eqnarray}
h_r = \frac{1}{2} + \frac{1}{8z}\left( 1- (z-u)^2\right) 
\log \left( \frac{|z-u+1|}{|z-u+1|} \right) \nonumber  \\
+ \frac{1}{8z}\left( 1- (z+u)^2\right) 
\log \left( \frac{|z+u+1|}{|z+u+1|} \right)\nonumber \\ \label{eq:re} \nonumber 
\end{eqnarray}
For partially degenerate plasmas, the susceptibility can be expressed as the superposition of the susceptibility of the completely degenerate gas based on the technique developed by Dharma-Wardana~\cite{wardana}. 

For the given density perturbation $\delta n$ as given in Eq.~(\ref{eq:density}), 
the 1-D BRS three-wave interaction  between the pump,  
the seed  and a Langmuir wave  
is described  by~\cite{McKinstrie}:
\begin{eqnarray}
\left( \frac{\partial }{\partial t} + v_p \frac{\partial}{\partial x} + \nu_1\right)A_p  = -ic_p A_s A_3  \nonumber \mathrm{,}\\
\left( \frac{\partial }{\partial t} + v_s \frac{\partial}{\partial x} + \nu_2\right)A_s  = -ic_s A_p A^*_3   \label{eq:2} \mathrm{,} \\
\left( \frac{\partial }{\partial t} + v_3 \frac{\partial}{\partial x} + \nu_3\right)A_3  = -ic_3 A_p A^*_s  
\nonumber \mathrm{,}
\end{eqnarray}
where $A_i= eE_{i}/m_e\omega_{i}c$  is 
the ratio of  the electron quiver velocity of the pump pulse ($i=p$)
and the seed pulse ($i=s$)  relative to the velocity of the light $c$,
 $A_3 = \delta n_e/n_e$ is the the Langmuir wave amplitude,
$\nu_1 $ ($\nu_2$) is the rate of the inverse bremsstrahlung  
of the pump (seed), 
$\nu_3$ is the plasmon decay rate, 
$ c_i = \omega_3^2/ 2 \omega_{i}$ for $i=p, s$, $c_3 = (ck_3)^2/2\omega_3$, and 
 $\omega_{3} \cong \omega_{pe} $  is  the plasmon  wave frequency.  
 The energy and momentum conservation of the three-wave interaction leads to  
$\omega_{3} = \omega_2 - \omega_{1} $ and 
$k_{3} = k_1 + k_{2} $,
where $k_3$ ($\omega_3$) is the plasmon wave vector (frequency).

Eq.~(\ref{eq:2}) is derived for the cold classical plasma by separating the fast varying resonance frequency from the slow varying pulse evolution~\cite{McKinstrie}.  For non-zero temperature classical plasma or degenerate plasmas, 
the first and second equations, which is just the simple Maxwell equation,  are unchanged but the third equation can be generalized from Eq.~(\ref{eq:density}) to 

\begin{equation}
\left( \frac{\partial }{\partial t} + v_3 \frac{\partial}{\partial x} + \nu_3\right)A_3  = -i\frac{2 c_3}{ \omega (\partial \epsilon / \partial\omega) } A_p A^*_s  \label{eq:3} 
\end{equation}
where we use $\epsilon=0$ so that $(4 \pi e^2 /k^2\chi_e) = -1$.  For the cold classical plasma with $\epsilon = 1 - \omega_{pe}^2 / \omega^2 $,  $\omega (\partial\epsilon / \partial \omega) = 2$ and Eq.~(\ref{eq:3}) is reduced to  Eq.~(\ref{eq:2}).  
The computation of $\omega (\partial \epsilon / \partial \omega) $ is the
most important in estimating  the quantum effect on the BRS.

\begin{figure}
\scalebox{0.3}{
\includegraphics[width=1.7\columnwidth, angle=270]{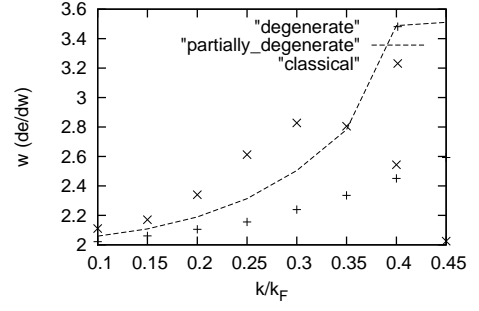}}
\caption{\label{fig2}
$-\omega (\partial \epsilon / \partial \omega) $ as a function of the wave vector $k/k_F$ for the classical plasmas, the partially degenerate plasma and completely degenerate plasma. The x-axis is $k / k_F $ and the y-axis is $\omega (\partial \epsilon / \partial \omega) $. 
In this example, $n_e = 10^{24} / \mathrm{cc} $, 
$T_e = 20 \ \mathrm{eV} $ and $E_F = 36\  \mathrm{eV} $.  
}
\end{figure} 

Using Eqs.~(\ref{eq:cla}) and (\ref{eq:quan}) and the integral transform provided by 
Dharma-Wardana~\cite{wardana},
we computed the derivative ($\omega (\partial \epsilon / \partial\omega$) for classical plasmas, partially degenerate plasms and 
degenerate plasma and illustrated them in Figs.~(\ref{fig1}) and (\ref{fig2}).
In Fig.~(\ref{fig1}), we plot
$\omega (\partial \epsilon / \partial \omega) $ 
at the Langmuir wave resonance ($\epsilon=0$) 
as a function of the wave vector $k/k_F$ when 
the electron density is   $n_e = 10^{24} / \mathrm{cc} $. 
We compute three cases. 
The first case is when the dielectric function is given classically
with the electron temperature $T_e = E_F$ (equation \ref{eq:cla}), 
the second case is when the plasma is completely degenerate and $\epsilon$ is given by Lindhard function~\cite{lindhard} (equation \ref{eq:quan}) and 
the third case is when the plasma is partially degenerate with the temperature $T_e = E_F$. 
When $k/k_F \ll 1$, all three cases are the same with 
 $\omega (\partial \epsilon / \partial \omega) \cong 2$.
As $k/k_F$ gets higher, 
 $\omega (\partial \epsilon / \partial \omega)  $ reaches the maximum at $k/k_F\cong 0.23 $ for classical plasmas, 
  $\omega (\partial \epsilon / \partial \omega)  $ reaches the maximum at $k/k_f \cong 0.4 $ for the partially degenerate plasma  and 
   $\omega (\partial \epsilon / \partial \omega)  $ is monotonically increasing for the degenerate plasma. 
For the classical plasma, when $k/k_F > 0.45 $, the plasmon no longer exists as $k \lambda_{de} \geq 0.5 $.
In the example provided, the Landau damping at $\epsilon = 0 $ is negligible.   
In general, the classical plasma overestimate (underestimate) the BRS coefficient given in the third equation of Eq.~(\ref{eq:3})  when $k/k_F$ is large (small). 
In Fig.~(\ref{fig2}), we plot  $\omega (\partial \epsilon / \partial \omega) $ 
for the same plasma in Fig.~(\ref{fig1}) but with the electron temperature $T_e = 20 \ \mathrm{eV}$. 
In this example, the Landau damping at $\epsilon = 0 $ is also negligible.  

So far, we  have applied the first order perturbation theory of the quantum particles in 
the presence of the x-ray laser field in order to obtain the density perturbation. 
In principle, we can extend  this calculation to the arbitrary order. 
Here, the author will adopt a simple approach of the Volkov state~\cite{volkov}
and compute the perturbation to the second order.
In this approach, the first order solution is given as 

\begin{equation} 
\psi^v_k  = \exp\left( - \frac{1}{\hbar} \int \left[ \frac{1}{2m_e}(i\hbar k+eA/c)^2\right] dt + ikx
\right) \label{eq:vol}
\mathrm{,}
\end{equation}
In the limit $k_1 + k_2 \ll k_F $, $\psi^v_k$ becomes the exact solution of Eq.~(\ref{eq:ham}). The convenient fact is that 

\begin{equation}
\frac{ \chi_e V_p } { \epsilon} = 
\int \left[ e^{ikx-i\omega t} \psi^v_{k_1}(x,t) \psi^{*v}_{k_1}(x,t) f_{k_1} \right]d^3 x dt d^3 k_1\mathrm{,}
\label{eq:use}  
\end{equation}
where $f_k $ is the electron distribution function. 
As in the case of the time-dependent perturbation theory, we express the quantum wave function as $\psi_k = \psi^v_k \psi^s_k$.  By substituting $\psi_k$ into   Eq.~(\ref{eq:ham}) and using  Eq.~(\ref{eq:vol}), we obtain the equation for $\psi^s_k$:

\begin{eqnarray} 
i\hbar\frac{\partial \psi^s_k}{\partial t}  
= &+&\left[ \frac{1}{2m_e} \left( i\hbar \nabla + \frac{e}{c} \mathbf{A}(x,t) +\frac{e}{c} \mathbf{B}(x,t)\right)^2\right]\psi^s_k \nonumber  \\    &-& \frac{1}{2m_e}\left[ \left( i\hbar k + \frac{e}{c} \mathbf{A}(x,t)\right)^2
\right] \psi^s_k  \mathrm{,} \label{eq:ham3}
\end{eqnarray}
where $(e\mathbf{B}/ c) = (m_e v^p_{osc} v^S_{osc} / v_{ph}) \cos((k_1 + k_2)x - (\omega_1 -\omega_2) t)$, $v_{ph} =  (\omega_1 -\omega_2)/ (k_1 + k_2) $, Eq.~(\ref{eq:pond}) is used and we ignore the fast varying part of the potential. By ignoring the fast varying part of the potential, 
 Eq.~(\ref{eq:ham3}) can be simplified to

\begin{equation} 
i\hbar\frac{\partial \psi^s_k}{\partial t}  
=  \frac{1}{2m_e}\left[ \left( i\hbar (k + \nabla) +\frac{e}{c} \mathbf{B}(x,t)\right)^2
   -  ( i\hbar k)^2\right] \psi^s_k
 \mathrm{,} \label{eq:ham4}
\end{equation}
From Eq.~(\ref{eq:ham4}), there are two additional contribution to the original pondermotive potential given in Eq.~(\ref{eq:pond}). The first one is the first order contribution given as 

\begin{equation} 
V^1_p = \frac{e\hbar k}{m_ec} \ \mathbf{B}(x,t) = 
   \frac{\hbar k }{m_ev_{ph}}  m_e v^p_{osc} v^S_{osc} \cos(k_3x - \omega_3 t) \mathrm{,} 
\label{eq:pond2}
\end{equation} 
where 
$k_3 = k_1 +k_2$, $\omega_3 = \omega_2 - \omega_1 $ and  
 we assume that $\hbar k \gg \hbar \nabla $. 
Note that $V^1_p / V_p \cong  (\hbar k/m_ev_{ph}) $. 
It is usually the case   $(\hbar k/m_ev_{ph}) \ll 1 $ in the BRS 
because the Langmuir wave whose phase velocity is greater than the electron thermal velocity needs to be excited in order to avoid the heavy Landau damping. 
The second order contribution to the pondermotive potential is 

\begin{equation} 
V^2_p  \cong m_e \left(\frac{  v^p_{osc} v^S_{osc} }{ v_{ph}^2}\right) 
       v^p_{osc} v^S_{osc} \cos(2k_3x - 2\omega_3 t) \mathrm{.}\label{eq:pond3}
 \end{equation}
The ratio of the second order pondermotive potential to the 
first order pondermotive potential is given as
$V^2_p / V_p \cong  (v^p_{osc} v^S_{osc}/  v_{ph}^2) $. 
The ratio is small when $v^p_{osc}/ v_{ph} < 1$. 
The second order contribution to the pondermotive potential $V^2_p$ is especially interesting 
in the sense that it arises even in the single particle motion; 
 it would be only possible via the Vlasov field in the classical mechanics. 
If $\epsilon(2k_3, 2\omega_3)=0$, then strong resonant Langmuir wave can be excited even if the original seed and pump x-ray does not satisfy the resonance condition.

In conclusion, 
we estimate the electron  degeneracy and diffraction effect 
on the backward Raman scattering of the x-ray lasers in dense plasmas. 
When $k / k_F<0.2$ ( $k / k_F>0.2$), the BRS in the partially degenerate and degenerate plasma is stronger (weaker) than  the classical plasma as shown in Figs.~(\ref{fig1}) and (\ref{fig2}) . 
We also derive one version of the quantum second order perturbation theory and show that 
the second harmonic of the pondermotive potential 
arises naturally even in a single particle motion. 
Our calculation in this paper is relevant when $T_e \leq E_F$ and 
$v_{osc} \leq v_{ph} $. 
In the BRS of dense plasmas, 
the inverse bremsstrahlung is an important factor to consider because the plamsa will heat 
up in a fast time scale.  For an example, when $n_e \cong 10^{24} \ /\mathrm{cc} $, 
the inverse bremsstrahlung could heat up the plasma temperature to the Fermi energy in a few thousands  Langmuir periods~\cite{sonbackward} if the x-ray lasers have the wave length of 1 nm and the intensity of $10^{18} \  \mathrm{W} / \mathrm{cm}^2 $.
Our calculation of the quantum deviation is only relevant 
in this initial phase of the BRS before the significant heating occurs.

The author uses the well-known dielectric function theory 
and the second order quantum mechanical perturbation theory; 
there is nothing new in the  methodologies.
But, the practical computation of the BRS coefficients, including the electron degeneracy and diffraction,   are presented for the first time. 
 The physical processes  in dense plasmas
are significantly different from  from the classical prediction, 
of which the studies are still rare.  
The analysis in this paper is one of the meaningful results in this effort.

\bibliography{tera2}

\end{document}